\documentclass[runningheads]{llncs}
\usepackage[T1]{fontenc}
\usepackage{graphicx}
\usepackage{multirow}
\usepackage{threeparttable}
\usepackage{comment}
\usepackage{booktabs} 
\usepackage{xcolor}
\usepackage{subfig}
\usepackage{url}
\usepackage{tabularx}
\usepackage{hyperref}
\usepackage{array}
\begin{document}
\title{What Does a Meow Mean?  In Search of Intuitively Understandable Communication by a Nonverbal Companion Robot}

\titlerunning{What Does a Meow Mean?}

%
\author{Vivienne Bihe Chi\inst{1} \and
Claudia B. Rébola\inst{2}
\and
Bertram F. Malle\inst{1} 
}

\authorrunning{V. B. Chi et al.}
%
\institute{Brown University, Providence RI 02912, USA 
\email{\{vivienne\_chi,bfmalle\}@brown.edu}
\\
\and
University of Cincinnati, Cincinnati OH 45221, USA\\
\email{rebolacb@ucmail.uc.edu}
}

\maketitle              
\begin{abstract}
Older adults living alone have a number of challenges, and robots can help with some of them---by providing reminders, initiating activity, or offering comfort. As part of developing a cat robot with limited assistive functions, we designed a set of nonverbal communication signals, both auditory (cat sounds) and visual (icons on a small display). To evaluate these signals we used a mixed-methods, user-centered approach.  After a pilot study, a focus group with older adults suggested revisions to the initial signal set. A large-sample online experiment then tested whether adults over the age of 65 could accurately infer the robot's communicative intentions. When both visual and auditory signals were present, accuracy was high. When visual signals were absent, accuracy often decreased; when auditory signals were absent, accuracy sometimes increased. So the auditory signals were less helpful, except when the robot conveyed strong sentiments (e.g., purring while being petted). 

\end{abstract}
\section{Introduction}

Nearly three in ten older adults reside alone in the U.S. \cite{Hemez_2024} 
and face  challenges from falling to accidental prescription overdose to loneliness \cite{Donovan2020}.  Robots may help alleviate these challenges, but older adults often encounter barriers when operating technology. Intended to simplify their lives, technology frequently causes anxiety and feelings of incompetence. Any assistive robot must have an intuitive and accessible design,   
enabling effortless interaction and facilitating adoption and long-term use. 
In such design, understandable communication is central.   

\begin{figure}[ht]
    \centering \includegraphics[height=4cm]{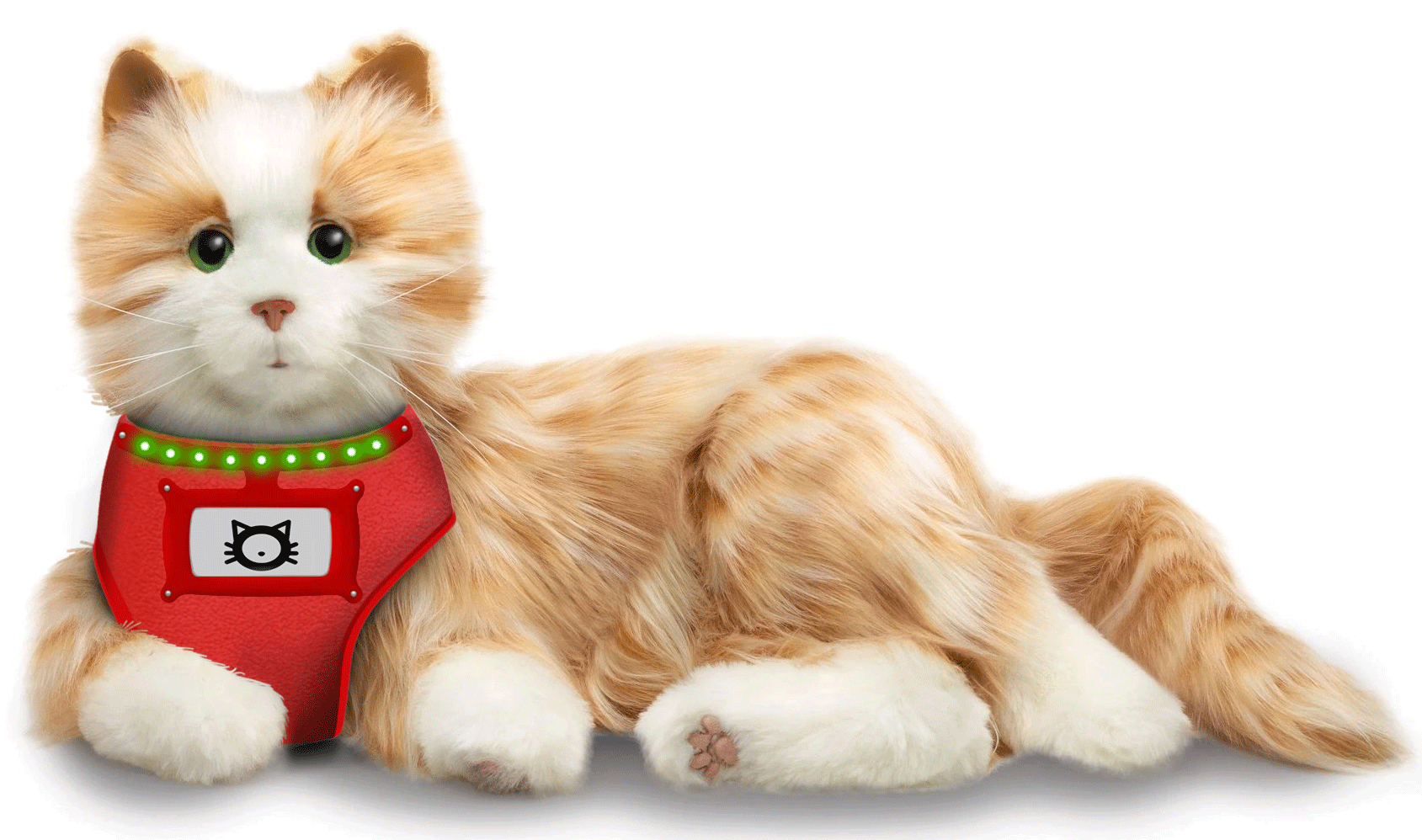}
      \caption{\textit{Ageless Innovation's} Joy For All cat, with a smart collar added.}
      \label{fig:cat_image}
    \end{figure}

We have been developing a zoomorphic robot (built on the \textit{Ageless Innovation} Joy For All animatronic cat \cite{JFA-cat}, Figure~\ref{fig:cat_image}), designed with limited assistive functions but intended to be affordable for broad deployment. In developing the communicative signals that the robot emits, we needed to confirm that any human user can easily understand these signals. We report on the development and evaluation of intuitively understandable communication signals expressed by this nonverbal robot. We used a mixed-methods, user-centered approach, including a focus group study and a large-sample online experiment with older adults.





\section{Related Work}

\subsection{Social Robots for Older Adults} 

The proportion of older adults is increasing globally, with those aged 65 and older projected to represent an even larger share of society in the near future \cite{us_census_bureau_aging_2020}. 
There are rising concerns about addressing their social and emotional needs, as many experience isolation and loneliness \cite{ausubel_older_2020}. Social robots have emerged to offer companionship \cite{berridge_companion_2023} and reduce loneliness by facilitating human-human interaction \cite{ling2022,lu_social_2023}. Social robots can serve a role in healthcare settings by providing mental stimulation, assisting with daily tasks, and supporting health monitoring \cite{civit_introducing_2024,Trainum2024}. For older adults with dementia, these robots can lower stress, help the person keep a schedule, and improve well-being \cite{deutsch_home_2019}. 



Many challenges remain in the integration of social robots into older adults' lives. Some healthy older adults feel uneasy about being led to believe these robots are real companions \cite{lu_social_2023,berridge_companion_2023}. Further, privacy concerns about their advanced sensory capabilities and the handling of personal data are common, as are worries that robots could reduce opportunities for  human-human interaction \cite{berridge_companion_2023,liu_can_2021,gonzalez-gonzalez_social_2021}. Addressing these concerns is essential if we want to ensure that social robots become meaningful and effective companions supporting older adults.

\subsection{Animal Companion Robots}

Companion robots may be particularly important in assisting older adults. They  don't clean and iron and are often stationary, but they engage the user through touch, sound, and reactive movement, which creates a comforting, stress-reducing presence \cite{moyle_care_2018,hudson_robotic_2020}.    
In addition to emotional comfort, companion robots can provide cognitive stimulation and even healthcare support. Some are designed to assist with memory exercises and daily routines \cite{moyle_care_2018,jeong_robotic_2023}, others can support compliance with medication schedules
or promote healthy behaviors \cite{Mappo,Lv2023}. 

Companion robots often mimic the appearance of familiar pets such as cats and dogs \cite{moyle_care_2018,Miro2018}, but they can also take on less conventional shapes, such as Paro the seal, PLEO the dinosaur, or Baby Whale \cite{moyle_care_2018,Lv2023,Cho2011}. Interaction complexity varies, from Paro's subtle movements to Aibo's dynamic behavior \cite{bradwell_companion_2019}.  
Older adults, especially those experiencing loneliness or cognitive decline, prefer familiar forms like robotic cats and dogs for their emotional comfort \cite{hudson_robotic_2020,thunberg_social_2021}. For older adults, robots' interactive movement and responsiveness to simple commands are more important than technical sophistication \cite{bradwell_companion_2019}.   

\subsection{Universal Design \& Accessibility}

Human-centered design aims to make companion robots functional and accessible for distinct user groups with different needs, abilities, and preferences. The principles of Universal Design, in particular, favor products that are usable by people of all abilities without requiring adaptations \cite{nanavati_design_2023,wong_perspectives_2024}. Thus, social robots should be accessible to older adults with sensory, cognitive, or mobility changes often associated with aging \cite{farage_design_2012}. A fundamental principle of accessibility is the use of redundant communication signals by delivering information through multiple channels \cite{Donini2024}, such as visual, auditory, and tactile cues \cite{neto_im_2024,horton_review_2017}. This redundancy allows robots to convey critical messages to users with diverse abilities.

\subsection{Nonverbal Communication}

Verbal communication abilities are ubiquitous in virtual agents and humanoid robots, but such abilities would be unusual, perhaps disturbing, when given to animal robots like cats and dogs.  Pet companion robots are therefore best equipped with nonverbal communication \cite{Duffy2003}. 
In human-human interactions, nonverbal cues of smiling, nodding, and gestures are highly influential in social interactions \cite{Kendon2004}, from job interviews \cite{Tescari2024-bp} to physician-patient communications \cite{Ambady2002}.
However, a smiling, nodding cat robot may be just as uncanny as a talking one \cite{ZoomorphicUncanny}. Some researchers have applied insights from ethology to design robot behaviors that resemble natural animal behavior \cite{lakatos_emotion_2014} and are therefore easily readable by human users \cite{darling_new_2021}. Indeed, older adults seem to expect and prefer zoomorphic robots with natural sounds and colors  \cite{Collins2024}. 

Selecting communication signals solely from natural origins, however, may limit a companion robot's design.
Natural and artificial signals can be combined, as in the case of Cozmo's ``emotion'' expressions that include shape of eyes, head position, forklift movements, and sounds. Such complex expressions take time to unfold, however, and are potentially ambiguous \cite{pelikan_are_2020}.  Additional artificial signals may include differently colored lights \cite{passler_bates_caring_2024} or icons on display screens that indicate mood \cite{Chirapornchai2021} or that provide reminders \cite{Lv2023}. 

Extensive work is available on sound design for robots \cite{robinson_intro}.
Some have argued that all robotic sounds should strive to convey affective qualities \cite{HugMisdariis}, and in practice, robotic sounds are often used to communicate emotions \cite{Yilmazyildiz2015,Wolfe2024}. However, people cannot easily read a robot's vocabulary of emotions (and many will have doubts about a robot \textit{having} emotions).  
In one study, sounds alone successfully conveyed an intended emotion only around 40\% of the time, whereas adding congruent multimodal signals increased  the success rate to as much as 70\%  \cite{Robinson2022}. Several researchers have emphasized the need for using  redundant multimodal signals for robots' effective emotion communication  \cite{loffler,Donini2024}. Recent assistive pet robots have implemented this principle, such as MAPPO (using displayed icons and barking)\cite{Mappo} and Aibo (using spontaneous movements and sounds) \cite{Maignan2024}.

\section{Robotic Cat Design}

\subsection{Capacities and Appearance}

From the start we aimed at an affordable, limited-function companion robot that would be suitable for adults living alone. Thus, the design goals for our robot focused on assisting the user in affective-cognitive tasks---e.g., providing reminders, connecting with loved ones, facilitating emotional comfort through physical touch. In a previous study \cite{malle_developing_2019}, we had asked older adults ($N$ = 179), as well as caregivers of older adults ($N$ = 105), to indicate how much they liked a dozen zoomorphic robots, including the Joy-for-All (JFA) dog \cite{JFA-dog} and cat \cite{JFA-cat}, as well as Paro, Aibo, Owl, Dragonbot, and others.  The two JFA animatronic animals were in first place in both samples (mirroring earlier results \cite{bradwell_companion_2019}). 



To implement multiple communication channels, we equipped the Joy-for-All cat model \cite{JFA-cat} with an ESP32 microcontroller for audio output and with a smart collar to display visual signals (e-paper display and LED lights), controlled by a Raspberry Pi. 
The Pi is powered by a battery recharged via a USB cable when the cat is placed on a ``bed'' connected to a normal power outlet. 


\subsection{Communication Channels}

We designed auditory signals natural for a cat (meow sounds and purring) and added visual signals (LEDs on the collar, icons on the e-ink display).  On the auditory channel, we initially used the off-the-shelf JFA meow sounds but, in response to user feedback, we designed more natural ones from actual cat sound recordings (for details, see Table \ref{tab:signalpackages} and \href{https://bit.ly/3QFkrp7}{audio files} at \url{https://bit.ly/3QFkrp7}). The LEDs can be programmed to show different flashing patterns, speeds, and colors. The e-ink display can be programmed to show detailed information, but we limited the signals to simple icons that are easily visible on the small screen (e.g., a pill, a power plug).  The specific signals and their combinations are described in Sect.\ref{Materials}.




\subsection{Overview}

We took a mixed-method approach to iteratively improve the robot cat's communicative signals.  
We first report on a pilot study (Sect.\ref{pilot study}) that tested the feasibility of an initial set of eight signal ``packages.''  Then we describe a focus group 
interview (Sect.\ref{focus group}) in which older adults discussed and provided feedback on a revised set of those packages. Finally, we report on a large-sample online experiment (Sect.\ref{online experiment}), in which we systematically tested revised communication signals across a wide age range.
In this experiment, preregistered at {\url{https://osf.io/eb3ht/overview}, 
we experimentally compared complete signal packages to incomplete ones (i.e., sound missing or icons missing) and collected extensive qualitative and quantitative information on people's comprehension of the signals and impressions of the robot. 

\section{Pilot Study}
\label{pilot study}

We developed eight use case scenarios centered on daily routines of older adults (e.g., reminder for an event, locating reading glasses) and designed eight initial signal packages (combinations of sound, LED, and icon) to fit the scenarios. In some scenarios, the robot initiated (e.g., medicine reminder, invitation to play); in others, it responded (reaction to petting, response to request for finding reading glasses). In addition, because the robot's battery level is essential for continued interaction, we tested three related communications (request for charging, affirmation that cable is plugged in, indication that charge is full). 
We then evaluated the interpretability of these signal packages in an online survey ($N$ = 105 recruited from Prolific). For each scenario, participants were shown a GIF of a cat wearing a smart collar with an icon on the display screen (see Fig. \ref{fig:cat_image}) as well as LED lights flashing in specific colors and patterns, accompanied by corresponding meowing sounds. For each signal, participants first provided an open-ended interpretation and then considered several potential interpretations provided by the researcher and ranked them in order of likelihood. We classified the open-ended responses for fit with the intended communicative message, and the majority of participants correctly inferred the message. Likewise, in all eight scenarios, participants predominantly considered the intended interpretation as the most likely (average rank < 1.7). Participants also rated (on a 0-10 scale) their ease of understanding the robot cat at 6.6, their liking of the robot at 7.0, and the robot's potential as a pleasant companion for older adults at 7.2.
Thus, the pilot study provided assurance that the designed robotic signals were reasonably comprehensible to a general audience and that the nonverbal communications positively influenced the users' perception of the companion robot.

\section{Focus Group Interview}
\label{focus group}


Four individuals (one male, three female) who resided in a senior care home in Cincinnati, Ohio participated in a 90-minute interview. All were over the age of 65. One participant had a severe hearing impairment, and her informal caregiver was the second participant. The third participant used a wheelchair for mobility and owned an older model of a companion pet robot. She had previously cared for a pet cat. The fourth participant was living with a pet dog.

In a hybrid format, an on-site interviewer directed the live discussion while another researcher was on video conference and took notes. To begin, the interviewer presented a picture of the cat robot on a computer screen, and participants discussed what they would like the robot to do. Then the researchers explained the cat robot's three main signal channels.
Next, participants responded to four robot-user interactions. In two of them, the robot initiated an interaction (requesting attention) or had a specific objective (medication reminder).  In the other two, the robot responded to the user's action with a sentiment (enjoyment of being petted) or affirmed a request.  
The cat's behavior was shown as a gif accompanied by sound, which could be replayed upon request. (We did not present a physical cat in order to prevent overengagement with its embodiment and physical feel.) 
For each scenario, the interviewer asked, ``What might the cat be trying to communicate?'', with potential follow-up questions such as, ``What do you think the sound indicates?'' 



In all four scenarios, participants indicated that the signals were easily noticeable. 
They were able to provide their own interpretations of the signals without hesitation (e.g.,``the cat wants to be rubbed and petted,'' ``the cat is trying to respond to my request''), and these interpretations were consistent with the intended messages. When prompted, participants referred to all three signal channels to support their understanding of the cat robot’s communicative intent.

Participants looked for emotional cues within the auditory signals. They recognized urgency and concern in two sounds that had been designed to express those sentiments. The meow sounds varied in repetition, intonation, and speed, and participants perceived an eagerness when the meowing was repeated. Two participants mentioned that the meowing sometimes appeared hyperactive or even  worried.  Participants identified the LED patterns as indicating urgency but raised important concerns about the flashing blue LED (as a reminder to take medicine). All participants immediately associated the blue light with a medical emergency. 
For other flashing patterns, participants diverged in their interpretations.
Participants found the visual channels most appealing (``so powerful, we want to see more of those!'').  Overall, the focus group's responses confirmed that our communication signal designs were on the right track. However, some of the meow sounds and some LED colors and patterns carried unintended associations.

\section{Experiment}
\label{online experiment}

With valuable feedback from the focus group in hand, we revised the communication signals, replacing the original synthetic audio clips with sounds sampled from actual cat recordings (purring, meowing) and changing some of the LED patterns. We then designed an online experiment to compare the individual and joint contributions of both sounds and icons to people's comprehension (LED patterns were held constant). 
We recruited a large sample of older adults representative of the target user population and 
collected their quantitative and qualitative  impressions of the robotic cat.

\subsection{Participants}

We recruited participants above the age of 65 ($N$ = 315, median age = 70) from Prolific. 205 identified as women, 106 identified as men, 1 preferred not to say. The median duration of the online study was 23.5 minutes, and participants were compensated at a rate of \$12/hr.

\vspace{-2mm}
\begin{table}[ht]
\centering
\setlength{\tabcolsep}{10pt}
\caption{Use case scenarios for the online experiment}
    \renewcommand{\arraystretch}{1.3}
\label{tab:scenarios}
\begin{tabular}{|l|l|l|l|}
\hline
\textbf{Direction} & \textbf{Purpose} & \textbf{Priority} & \textbf{Use case scenario} \\
\hline

\multirow{4}{*}{Initiate} 
& \multirow{2}{*}{Social} 
& High & Invites to play \\ \cline{3-4}
& & Low & Checks in \\ \cline{2-4}

& \multirow{2}{*}{Functional} 
& High & Medicine reminder \\ \cline{3-4}
& & Low & Time to hydrate \\
\hline

\multirow{4}{*}{Respond} 
& \multirow{2}{*}{Social} 
& High & Being petted \\ \cline{3-4}
& & Low & Being picked up \\ \cline{2-4}

& \multirow{2}{*}{Functional} 
& High & Doorbell \\ \cline{3-4}
& & Low & Lost reading glasses \\
\hline

Emergency &  & & Requests charging \\
\hline
\end{tabular}
\end{table}

\vspace{-2mm}
\subsection{Materials}
\label{Materials}
We organized eight plausible home interaction scenarios as a combination of three binary features (see Table \ref{tab:scenarios}): Direction (robot initiates or replies), Purpose (communicative goal is social or functional), and Priority (act is urgent or not). In addition to this 2 x 2 x 2 design, we added a ninth scenario (the ``emergency" of the robot needing its battery recharged) because it was well recognizable in pilot studies (over 70\%) and thus served as a benchmark.  Table \ref{tab:signalpackages} shows the detailed signal packages. 

         
      
         
         
      
        

\subsection{Design}

We randomly assigned participants to three signal conditions: (1) full signal (display + sound), (2) sound without display, and (3) display without sound. The latter two conditions simulate the user experience of individuals with visual and hearing impairments, respectively. Each group of participants saw all nine scenarios in randomized order. The LED colors and patterns were always included in the signal packages because they served to attract attention rather than conveying specific content. Participants then worked through the same nine scenarios a second time so we could assess how much they learned (without feedback) merely from seeing the multiple options of what the robot \textit{could} communicate. 

    
\subsection{Procedure}

After participants provided consent, we conducted an audio check to ensure that participants could hear the auditory stimuli. Then they saw a picture of the cat robot equipped with the smart collar harness (Figure \ref{fig:cat_image}), introduced as a robot that ``will assist older adults with basic tasks of daily living.'' We invited participants to imagine having this robot cat in their homes and to anticipate its attempts to communicate in various scenarios. They were briefed on the available signals (sound, display, LED) in accordance with their assigned condition.

\newcommand{\rowbox}[3]{%
\fbox{%
\begin{tabularx}{0.72\linewidth}{X|X|X}
#1 & #2 & #3
\end{tabularx}}}

\begin{table}[h]
\centering

\caption{Signal packages for each use case scenario in the main experiment}
\vspace{2mm}
    \renewcommand{\arraystretch}{0.55}
\label{tab:signalpackages}
\begin{tabular}{p{3cm} l}

\textbf{Scenario} & \textbf{Icon \hspace{2.15cm} Audio \hspace{1.8cm} LED} \\[4pt]

Invites to play &
\rowbox{ball of yarn}{ curious meow $\times 3$}{yellow, flashing } \\[10pt]

Checks in &
\rowbox{cat paw up}{curious meow $\times 1$}{yellow, closing } \\[10pt]

Medicine reminder &
\rowbox{pills}{attention meow $\times 3$}{red, flashing } \\[10pt]

Time to hydrate &
\rowbox{glass of water}{attention meow $\times 1$}{red, closing } \\[10pt]

Being petted &
\rowbox{relieved cat face}{purr $\times 3$}{purple, flashing } \\[10pt]

Being picked up &
\rowbox{smiling cat face}{purr $\times 1$}{purple, chase } \\[10pt]

Doorbell &
\rowbox{spinning wheel}{affirming meow $\times 3$}{green, blinking} \\[10pt]

Lost reading glasses &
\rowbox{crossed out circle}{affirming meow $\times 1$}{green, closing } \\[10pt]

Requests charging &
\rowbox{battery and plug}{urgent meow $\times 3$}{blue, flashing } \\

\end{tabular}

\end{table}

\vspace{-4mm}

For each scenario, a brief narrative established the context (e.g., ``Imagine you are petting your cat. Now you see your cat does this: ''). (For contexts of all nine scenarios, see 
Supplementary Materials [SM], Sect. 1, {\url{http://bit.ly/4uwnyz9}.) Next, the communicative act appeared, depicting the cat robot emitting auditory and/or visual signals. Participants were asked, ``What might the cat be trying to communicate?'' and provided an open-ended interpretation. On the next page, six potential interpretations appeared in randomized order: the intended message (e.g., ``C'mon, let's play!'') and five competitors of varying plausibility, taken from other scenarios (e.g., ``Oh, hello there,'' ``Don’t forget to take your pills!'', ... ``I need to be charged up.''). Participants  indicated, for each option, how likely it was that the cat robot intended to communicate that message, using a slider scale ranging from -100 (Extremely Unlikely) to +100 (Extremely Likely).

After evaluating all nine scenarios in a first round, participants reported their estimates of how often (across the nine scenarios) they felt they were informed by each of the available signals, their ease of understanding (``How easy was it to understand the robot cat in these scenarios?''), the robot's likability (``After observing this robot cat in these scenarios, how much do you like it?''), their excitement for having such a robot cat around for themselves. Responses were recorded on a scale ranging from 0 (not at all) to 10 (very much).

Subsequently, participants were asked to ``imagine another day with the robot cat,'' where they experienced the nine scenarios in a second round. This time they  evaluated only how likely each of the six potential interpretations of the cat's communication were.  At the end of the second round, they again reported on their reliance on the available signals, their ease of understanding, the cat's likability, and their excitement for owning this robot cat.


\begin{table}
\centering
\begin{threeparttable}
\setlength{\tabcolsep}{7pt}
\caption{Accuracy of open-ended interpretations of robot communicative acts}
    \renewcommand{\arraystretch}{1.3}
\label{tab:accuracy}\begin{tabular}{llccc}
\toprule
Scenario & Direction | Purpose | Priority 
&    Full    & \shortstack{No\\sound} & \shortstack{No\\display}  \\
\midrule
Invites to play       &  Initiate | Social | High  & 71\%  & 80\%  & 52\% \\
Checks in             &  Initiate | Social | Low   & 40\%  & 55\%  & 22\% \\
Medicine reminder     &  Initiate | Functional | High  & 62\%  & 85\%  & 41\% \\
Time to hydrate   &  Initiate | Functional | Low   & 47\%  & 67\%  & 20\% \\
Being petted   &  Respond | Social | High  & 83\%  & 79\%  & 88\% \\
Being picked up    &  Respond | Social | Low   & 7\%  & 13\%  & 6\% \\
Doorbell       &  Respond | Functional | High  & 92\%  & 91\%  & 96\% \\
Lost glasses    &  Respond | Functional | Low   & 48\%  & 52\%  & 24\% \\
Requests charging     &  Emergency  & 79\%  & 70\%  & 49\% \\

\textbf{Median} & &  \textbf{62\%} & \textbf{70\%} & \textbf{41\%} \\
\bottomrule
\end{tabular}

\begin{tablenotes}
\footnotesize
\item \textit{Note.} Percentages are the proportion of people in that cell whose verbal interpretation passed the lenient accuracy criterion.
\end{tablenotes}

\end{threeparttable}
\end{table}

\subsection{Results}

\subsubsection{Open-ended accuracy}

Upon seeing each communicative act, participants provided their open-ended interpretations of the act. To form initial coding categories we used a bottom-up approach, applying cluster analysis to sentence embeddings of all 2,836 entries and then refining the categories (e.g., \textit{wants attention; needs charging}). Two coders then classified all entries into these categories (overall $\kappa$ = 0.90) and resolved disagreements by discussion.  (For details of the procedure, coding categories, and reliability, see \href{https://osf.io/5e389/files/vyhfw?view_only=78466639774f4600ad15346d81b0d197}{SM}, Sect. 2) 

By consensus, the coders  determined to what degree each coding category matched the intended communicative message for a given scenario. We assigned three kinds of scores: strict accuracy (yes, no), liberal accuracy (yes, no), and feature match (0-3 score). We report here the liberal accuracy results (interpretations that are either fully accurate or clearly defensible but not exactly as intended) and report on the other two scores in the \href{https://osf.io/5e389/files/vyhfw?view_only=78466639774f4600ad15346d81b0d197}{SM}, Sect. 2.3.  

Table \ref{tab:accuracy} shows substantial accuracy in many scenarios. People struggled with one scenario (\textit{Being picked up}), where they interpreted the intended signal of ``Hello'' (with a waving-paw icon) consistently as enjoyment (plausible but clearly not accurate).  We tested the impact of signal condition in the eight primary scenarios with a generalized mixed-effects model: accuracy (0, 1) of each interpretation, predicted by fixed effects of signal condition (full signal, no sound, no display), the three features of direction, purpose, and priority, and their interactions, as well as a random intercept for participant. (Controlling for age did not change the results.)  With only cat sounds and LED, accuracy dropped considerably relative to the full condition ($z = -3.36$, $p < .001$).  By contrast, with only icons and LEDs (omitting sounds), accuracy improved ($z = 2.55$, $p = .011$). 

The features of the communicative acts (Direction, Purpose, Priority)  substantially influenced  interpretation accuracy.  High-priority messages (with flashing  LED patterns and repeated sounds) were far easier to interpret (84\%) than low-priority messages (27\%), $z = -20.38$, $p < .001$. Also, communicative acts with functional purposes (68\%) were easier to interpret than those with social purposes (47\%), $z = 7.27$, $p < .001$. Importantly, effects of signal conditions were much weaker for acts of Responding than acts of Initiating. The drop in accuracy for no-display was only 1\% in Respond, relative to 25\% in Initiate ($z = -3.42$, $p < .001$), and the increase in accuracy for no-sound was only 4\% in Respond, relative to 21\% in Initiate ($z = 2.92$, $p = .004$).  Thus, redundancy of signals was successful when the robot's communication responded to the user, but when the robot initiated, icons were essential for accurate interpretation, and sounds were somewhat interfering.   

\subsubsection{Accuracy measured by likelihood judgments}

What makes open-ended signal interpretations so difficult is that the range of possible communicative acts is unknown.  In everyday settings, a user may know that range---what their robot \textit{might }communicate.  We mimicked this situation by offering, after the open-ended interpretations, six potential meanings of each signal: the intended one and five alternatives. We asked participants to rate each option's likelihood  of being the robot's intended message (measured on a -100 to +100 scale). We conducted a preregistered mixed between-within ANOVA on \textit{relative likelihoods} as a measure of accuracy: the person's rating of the intended option minus the average of the alternatives ($M_{\mathrm{rel}}$). (For an alternative analysis, see \href{https://osf.io/5e389/files/vyhfw?view_only=78466639774f4600ad15346d81b0d197}{SM}, Sect. 3.) Overall, the intended option was seen as far more likely than the other options ($M_{\mathrm{rel}}$ = 84.2), and the  patterns of variation generally mirrored those of the open-ended data (see Figure \ref{fig:likelihood}).  
Specifically, compared to the full condition ($M_{\mathrm{rel}}$  = 86.3), accuracy decreased when the visual signal was missing ($M_{\mathrm{rel}}$  = 57.9) but increased when the sound was missing ($M_{\mathrm{rel}}$  = 108.4), both $t(1,312) > 3.8$, $p < .001$.  However, these variations of signal channel were less than half the strength in Respond communications compared to Initiate communications, $F(2,312) = 35.5$, $p < .001$. Moreover, they were also half the strength for Social compared to Functional communications. Thus, as Fig. \ref{fig:likelihood} shows, redundancy was successful when the robot responded to the user in a social scenario.  In addition, as in the open-ended data, high priority messages elicited considerably higher accuracy  ($M  = 105.4$) than low-priority messages ($M  = 63.0$), $F(2,312) = 261.8$, $p < .001$, and that was particularly true for messages with social purpose, $F(2,312) = 80.5$, $p < .001$.  

\begin{figure}[h]
 \vspace{-0.5em} 
        \centering
        \includegraphics[height=5.4cm]{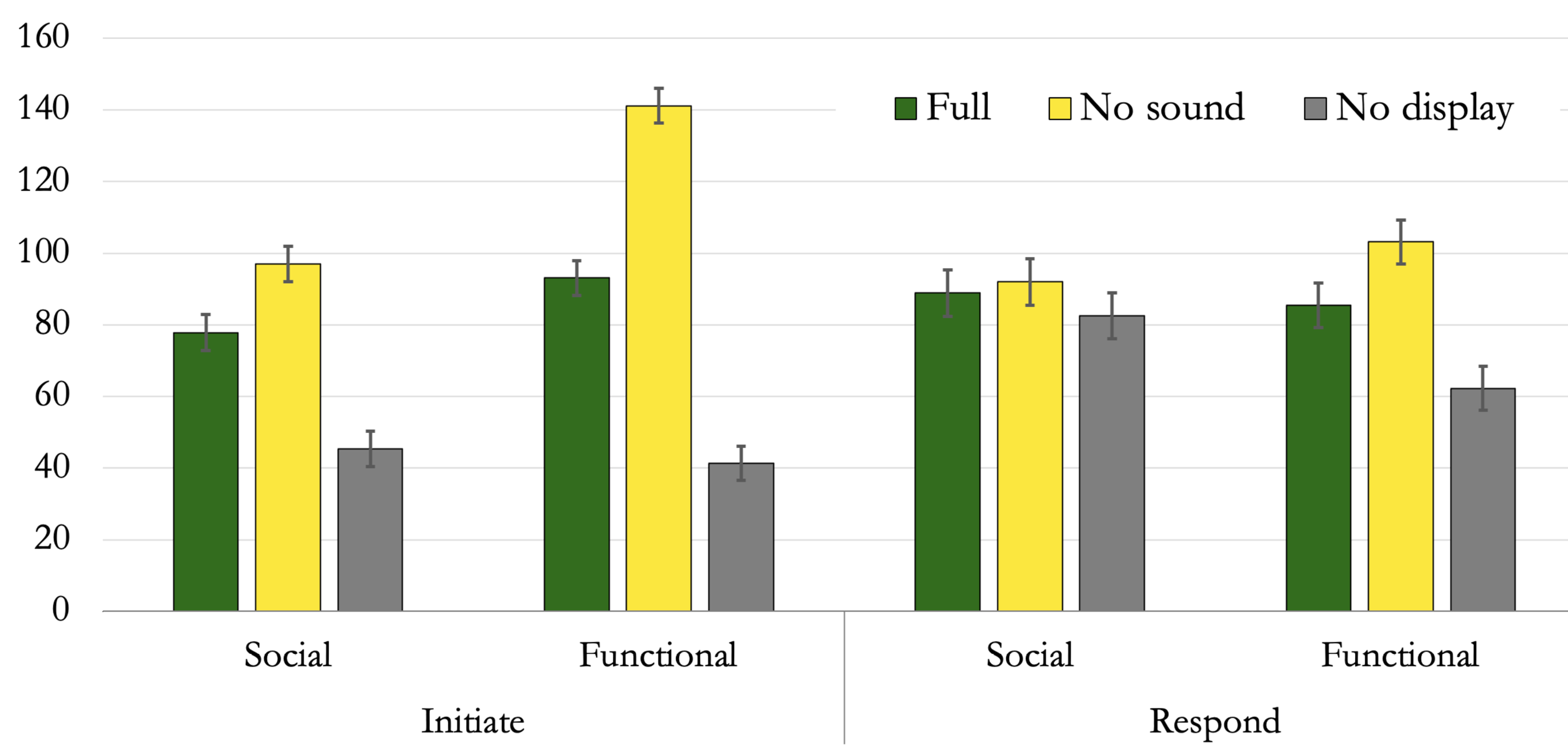}
        \caption{A measure of numeric accuracy: Rated likelihood of the correct option minus the average of the alternatives (range 0-200)}
        \label{fig:likelihood}
 
\end{figure}

\vspace{-3mm} 
   
\subsubsection{Improvement of accuracy with practice}

People's accuracy improved from the first to the second round of encountering and interpreting the signals. In round 2, their relative likelihood judgments ($M_{\mathrm{rel}}$) improved by 24 pts overall, $F(1,309) = 184.8$, $p < .001$. However, the improvement was more modest for the no-display condition (11 pts) than  the no-sound condition (25 pts) and especially the full-signal condition (35 pts), $F(2,309) = 15.2$, $p < .001$.  Interestingly, participants in the no-display condition improved on the Respond acts (21 pts)  but barely on the Initiate acts (2 pts), whereas people in the other conditions improved about equally in Respond and Initiate.  Thus, the most difficult signal condition remained difficult, even after added familiarity. Likewise, we also see that the overall more challenging \textit{social} communicative acts improved less (18 pts) than the functional acts (30 pts).   

\subsubsection{User acceptance}


People's ratings of how much they liked the robot and how excited they would be to have this robot around were highly correlated ($r > .86)$, so we averaged them and call that score \textit{acceptance}. The level of acceptance was modest overall ($M = 5.9$) and significantly lower than perceived safety  ($M = 7.7$), $t(314) = 11.1, p < .001$.  However, acceptance increased from the end of the first round  ($M = 5.77$) to the end of the second round ($M = 6.04$), $t(314) = 3.52, p < .001$. When we predicted acceptance at the end of the first round from  signal condition, participants in the three conditions did not differ, $t$s > 1.4, $p$s > .17.  When we predicted acceptance after the second round, the no-display condition did show a significantly lower level of acceptance ($M = 5.4$) than  the full condition ($M = 6.4$),  $t(312) = 2.21, p = .028$. (The no-sound and full condition did not differ at either time point.)  But this effect of the no-display  condition emerged only because the full condition elicited  increased  acceptance (from 5.9 to 6.4)  whereas the no-display condition held steady at 5.4. Moreover, the effect of the no-display condition on  acceptance was much weaker ($d = 0.31$) than its effect on perceived ease of understanding ($d = 1.18$). The latter increased slightly for no-display  participants (from 5.0 to 5.6) but much more so for full-signal (from 6.4 to 8.2) and no-sound participants (from 7.1 to 8.4). All in all, we can say that even when our older participants struggled with the no-display robot (were less accurate and experienced it as more challenging to understand), they accepted the robot barely any less than the other groups.   Admittedly, acceptance overall was somewhat tepid.  

\section{General Discussion }
We presented the design and evaluation of a nonverbally communicating companion robot.  On the basis of an earlier study of older adults' preferences for certain robot forms \cite{malle_developing_2019}, we developed a cat robot with limited capacities but multi-modal communication signals for a range of use cases.  Over the course of a pilot study and a focus group interview, we refined the packages of signals (combinations of icons, sounds, LED patterns) that might best represent certain communicative intentions.  In a large online study with more than 300 participants over the age of 65, we assessed whether participants can interpret these signal packages without any prior instructions. We also compared  visual and auditory signal channels and their possible (beneficial) redundancy.   Our studies yielded a number of insights on the topic of multi-modal communication signals and also suggest areas of concern, limitations, and directions for further research.      

\subsection{Insights Gained}
The most important insight was that the over-65 participants were often able to correctly infer, without any manuals or training,  the robot's communicative messages.  This success emerged both in the open-ended and numerical accuracy measures and further increased in a second practice round. This result affirms the hope that human-centered design can enable users to have spontaneous, successful interactions with nonverbal robot companions and that these interactions are apt to improve over time.
 
Second,  most people correctly inferred the intended messages when all channels (visual icons, natural cat sounds, and LEDs) were in place (median = 62\%). Accuracy was even higher when sounds were absent but icons were available (median = 70\%), and significantly lower when the visual display was blank but sounds were available (median = 41\%). Thus, signals were not truly redundant, as visual icons were more effective than sounds. However, the redundancy of channels varied by message type (\textit{cf.} \cite{loffler}).  When the robot initiated an action (and the user did not expect it), the icons were needed to infer the robot's intention and specific references to the physical world. When the robot responded to a user-initiated action (especially a social one), sounds relayed sentiments and were able to make up for the absence of icons. This observation aligns with the results from the focus group interview, which indicated that individuals relied on their interactions with real animals to discern latent information such as urgency and emotional cues from animal sounds. Thus, future designs could tailor multi-modal nonverbal signals to specific types of communication.   More complex intentions and references to objects may require specific visual icons; social and emotive communication can succeed with any signal channel.  

A further insight was that, even though the no-display robot elicited lower accuracy and was more difficult to  understand, participants accepted that robot cat no less after the first round of interaction.  After the second round, people substantially improved in accuracy in the other two conditions (icon only and full) and experienced the interpretation task as far easier than in the first round.  Improvements for the no-display condition were more modest, and their acceptance did not increase as it did in the other conditions.   Overall, however, both accuracy and perceived ease of understanding increased in the second round for all participants, suggesting that they were motivated to engage with the robot, made sense of it, and found that task rewarding.   This sustained engagement and improved accuracy of interpretation is particularly promising given that our  participants were 65 to 85 years old.  The levels of accuracy they achieved were no less than ones we observed in earlier studies with younger populations.


\subsection{Limitations and Future Work}

Despite our attempts to make communicative acts recognizable, we failed in two scenarios.  In one, participants had to imagine picking up the robot, upon which the robot emoted something like  ``Hello there.''  Participants predominantly read this signal as an expression of enjoyment, not as a greeting (and in hindsight we agree with this interpretation).  In the second case, participants had to imagine asking the robot whether it heard the doorbell, and they interpreted its response---intended as  ``Processing....let me check''---as a more specific statement about someone being (or not) at the door.  These cases illustrate that, as researchers and designers, we must always test communicative signals with our stakeholder population, even when the signals' meanings seem clear to us.  

The communication scenarios in the current studies are still only a subset of the relevant interactions that older adults might have with robot companions. Future research must develop a broader vocabulary for additional interactions and robot functionalities.  However, we caution against expanding this vocabulary too far, because having to track a large number of signals may present cognitive challenges that hinder adoption.

The robot prototype we are working with was designed to have limited capacities but may still be too limited for people to value its presence and sustain long-term user interest \cite{passler_bates_caring_2024}.  Some countermeasures could include providing the robot with learning capabilities (e.g., for user preferences) or adding more task-oriented functions. Which ones are most useful (e.g., connecting to a smart phone, fall detection) must be studied in future investigations.  Amid them, the functions of providing comfort and encouraging the adult to ``take care'' of the robot (at least by petting or charging it) have measurable benefits \cite{Tost2024} that may be most important to older adults, especially those with cognitive decline.

Even though we designed multi-modal communication signals, they are still limited.  Additional tactile signals may be useful for people with complete loss of sight \cite{Voysey2023}. The LED colors and patterns were mainly intended to be attention-alerting, but without longer-term user evaluations, we do not know whether people find them useful or distracting \cite{passler_bates_caring_2024}.  Finally, the current stationary robot has very few gestures, whereas some animal-shaped robots (e.g., for children with ASD) give emotion responses via gesture\cite{Burns2021}.  In the future, the LED patterns could stand in for deictic gestures as navigational cues when helping the user relocate lost objects \cite{Angelopoulos2022}.

The results of this controlled, online experiment cannot speak to the complexities of real-world attention, enjoyment, or fully contextualized interaction. We did our best to mimic some real-world features: sample systematically from potential scenarios, provide context descriptions before each signal, introduce signals in random order, and create repeated exposure to enable practice.  We gained valuable insights into adults’ impressive ability to interpret the robot’s communication and the conditions under which visual and sound signals may be redundant.  Now we have a foundation for using such a robot in a real-world study.


\bibliographystyle{splncs04}
\bibliography{Main_reference_list}

\end{document}